\documentclass[a4paper,12pt,final]{article}

\usepackage[latin1]{inputenc}

\usepackage[in,plain]{fullpage}

\usepackage{amsmath,amssymb,bm,graphicx,mathptmx,cite}
\usepackage[small,bf]{caption}

\usepackage[colorlinks,linkcolor=blue,citecolor=blue,urlcolor=blue]{hyperref}
\makeatletter
\renewcommand\@makefntext[1]{\leftskip=0.0em\hskip-0.5em\@makefnmark{#1}}
\makeatother
\newcommand{\abs}[1]{\ensuremath{\lvert\mkern1mu{#1}\mkern1mu\rvert}}

\begin{document}

\renewcommand{\thefootnote}{\fnsymbol{footnote}}

\begin{titlepage}

\thispagestyle{empty}

\begin{center}

${}$

{\Large\textbf{A probabilistic cellular automata model for the dynamics of \\ a population driven by logistic growth and weak Allee effect}}

\vspace{6ex}

{\large\textbf{J. Ricardo G. Mendon\c{c}a}}\href{https://orcid.org/0000-0002-5516-0568}{\includegraphics[scale=0.35]{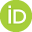}}\footnote{Email: \href{mailto:jricardo@usp.br}{\texttt{jricardo@usp.br}}}

\vspace{2ex}

\textit{\mbox{Escola de Artes, Ci\^{e}ncias e Humanidades, Universidade de S\~{a}o Paulo} \\ \mbox{Rua Arlindo Bettio 1000, Vila Guaraciba, 03828-000 S\~{a}o Paulo, SP, Brazil}}

\vspace{4ex}

{\large\textbf{Abstract}}

\vspace{2ex}

\parbox{5.25in}
{We propose and investigate a one-parameter probabilistic mixture of one-dimensional elementary cellular automata under the guise of a model for the dynamics of a single-species unstructured population with nonoverlapping generations in which individuals have smaller probability of reproducing and surviving in a crowded neighbourhood but also suffer from isolation and dispersal. Remarkably, the first-order mean field approximation to the dynamics of the model yields a cubic map containing terms representing both logistic and weak Allee effects. The model has a single absorbing state devoid of individuals, but depending on the reproduction and survival probabilities can achieve a stable population. We determine the critical probability separating these two phases and find that the phase transition between them is in the directed percolation universality class of critical behaviour.

\vspace{2ex}

{\noindent}\textbf{Keywords}: \mbox{population dynamics}~$\cdot$ \mbox{Allee effect}~$\cdot$ \mbox{mixed cellular automata}~$\cdot$ \mbox{diploid cellular automata}~$\cdot$ \mbox{cubic map}~$\cdot$ \mbox{directed percolation}

\vspace{2ex}

{\noindent}\textbf{PACS}: 02.50.-r~$\cdot$ 64.60.ah~$\cdot$ 87.23.Cc


\vspace{2ex}

{\noindent}\textbf{MSC 2010}: 92D25~$\cdot$ 82C22~$\cdot$ 60K35


\vspace{2ex}
{\noindent}\textbf{Journal ref.:} \href{https://doi.org/10.1088/1751-8121/aab165}{\textit{J.\ Phys.\ A:\ Math.\ Gen.}\ \textbf{51}~(14), 145601 [12~pp.] (2018)}}

\end{center}

\vfill

${}$

\end{titlepage}

\renewcommand{\thefootnote}{\arabic{footnote}}
\setcounter{footnote}{0}


\section{\label{intro}Introduction}

Cellular automata (CA) are discrete-space, discrete-time, discrete-state deterministic dynamical systems introduced in the 1940s as model systems for simple self-reproducing, self-repairing organisms and logical devices \cite{JvN1966,Wolfram94,Boccara10,Adamatzky13}. If the dynamics of the CA depends on a random variable, the CA becomes a probabilistic CA (PCA). PCA were introduced mainly by the Russian school of stochastic processes in its study of noisy, unreliable mathematical neurons, but also in relation with deep questions in the theory of Markov processes and statistical mechanics \cite{Russkiye90,Bennett85,Gacs01}.

In mathematical ecology, CA and PCA can be used to model the dynamics of populations that have discrete breeding seasons with nonoverlapping generations, like several species of insects and annual plants that can be described by difference equations, as well as in the study of spatial processes in ecosystems and land-use change \cite{Hogeweg88,Caswell93,Molofsky94,%
Durrett94,RandWilson95,Kareiva1997,Baltzer98,Silvertown92,Turner01,%
Wagner97,Benenson04}. One of the most interesting effects that may occur in the dynamics of a population is the Allee effect, according to which at low population densities reproduction and survival of individuals may decline, challenging the view that individual fitness must be higher at low densities because of lower intraspecific competition \cite{gotelli,allee,stephens}. Possible causes for the onset of the Allee effect in a population are mate limitation and debilitated cooperative defense, amongst others \cite{gascoigne,kramer,boukal,hastings}.

In this paper we propose and investigate a one-parameter probabilistic mixture of one-dimensional elementary ($0$-$1$) cellular automata under the guise of a model for the dynamics of a single-species unstructured population with nonoverlapping generations in which individuals have smaller probability of reproducing and surviving in a crowded neighbourhood but also suffer from isolation and dispersal. We found that a simple rationale for choosing the microscopic transitions that enter the PCA yields in the first-order mean field approximation to the dynamics of the PCA a cubic map containing terms representing both logistic limitation to growth and Allee effects. The model has an absorbing state devoid of individuals, but depending on the reproduction and survival probabilities can achieve a stable population. The two phases are separated by a second-order phase transition in the directed percolation universality class of critical behaviour.

The paper is organised as follows. In Section~\ref{sec:pca} we briefly review the PCA formalism and the single-cell mean field approximation to the dynamics of PCA. In Section~\ref{sec:mixed} we describe the Allee effect and discuss the rationale for the choice of the microscopic transition probabilities that renders a one-parameter PCA embodying both the logistic and the (weak) Allee effects. The mean field analysis of the model is presented in Section~\ref{sec:mfield}. In Section~\ref{sec:critical} we determine the critical behaviour of the PCA by Monte Carlo simulations and finite-size scaling analysis and briefly discuss its meaning in the general context of the model. Finally, in Section~\ref{sec:summary} we summarise our results and indicate some directions for further investigation.


\section{\label{sec:pca}PCA dynamics and mean field approximation}

A one-dimensional, two-state PCA \cite{Wolfram94,Boccara10,Adamatzky13,%
Russkiye90} is defined by an array of cells arranged in a one-dimensional lattice $\Lambda=\{1, 2, \dots, L\} \subset \mathbb{Z}$ of total length $L$, usually under periodic boundary condition ($L+i \equiv i$, $i=1, \dots, L$), with each cell in one of two possible states, say, $x_{i}=0$ or $1$, $i=1,\dots,L$. The state of the PCA at instant $t=0$, $1, \dots$ is given by $\bm{x}^{t}=(x_{1}^{t},\,x_{2}^{t},\,\dots,\,x_{L}^{t}) \in \Omega = \{0,1\}^{\Lambda}$. The probability $P_{t}(\bm{x})$ of observing the PCA in state $\bm{x}$ at instant $t$, given an initial distribution $P_{0}(\bm{x})$, is given by
\begin{equation}
\label{eq:markov}
P_{t+1}(\bm{x}') = \sum_{\bm{x}\, \in\, \Omega}\Phi(\bm{x}'\,|\,\bm{x})P_{t}(\bm{x}),
\end{equation}
where $0 \leq \Phi(\bm{x}'\,|\,\bm{x}) \leq 1$ is the conditional probability for the transition $\bm{x} \to \bm{x}'$ to occur in one time step. The rules that map the state of $x_{i}^{t}$ into the new state $x_{i}^{t+1}$ depend only on a finite neighbourhood of $x_{i}^{t}$. In this work the neighbourhood is given by the $i$-th cell itself together with its two nearest-neighbors $i \pm 1$, and since the cells of the PCA are updated simultaneously and independently we have that
\begin{equation}
\label{eq:probs}
\Phi(\bm{x}'\,|\,\bm{x}) =
\prod_{i=1}^{L}\phi(x_{i}'\,|\,x_{i-1},\,x_{i},\,x_{i+1}).
\end{equation}

From (\ref{eq:markov}) and (\ref{eq:probs}), it is easy to show that the dynamics of the marginal probability distribution $P_{t+1}(x)$ of observing a cell in state $x$ at instant $t$ (equivalently, the instantaneous density of cells in state $x$) obeys
\begin{equation}
\label{eq:p1}
P_{t+1}(x_{i}') =
\sum_{x_{i-1},\,x_{i},\,x_{i+1}}\phi(x_{i}'\,|\,x_{i-1},\,x_{i},\,x_{i+1})\,
P_{t}(x_{i-1},\,x_{i},\,x_{i+1}).
\end{equation}
From this equation we see that the determination of $P_{t}(x_{i})$ depends on the knowledge of the probabilities $P_{t}(x_{i-1},\,x_{i},\,x_{i+1})$, which in turn depend on $P_{t}(x_{i-2},\,x_{i-1},\,x_{i},\,x_{i+1},\,x_{i+2})$ and so on, constituting a full many-body problem that in general cannot be solved exactly. The simplest approach to get a closed set of equations from (\ref{eq:p1}) is to approximate
\begin{equation}
\label{eq:mf}
P_{t}(x_{i-1},\,x_{i},\,x_{i+1}) \approx
P_{t}(x_{i-1})\,P_{t}(x_{i})\,P_{t}(x_{i+1}).
\end{equation}
This is the single-cell mean field approximation, which assumes probabilistic independence between the cells. Higher order approximations involving pairs, triplets or more cells provide increasingly better approximate descriptions of the dynamics of the PCA \cite{local,DickMarro99}, but we shall limit ourselves to the simple approximation (\ref{eq:mf}).


\section{\label{sec:mixed}A mixed PCA inspired by population dynamics}

The model we are interested in was conceived as a toy model for the dynamics of a single-species unstructured population in which individuals suffer from local overcrowding but also from isolation and dispersal. In classical population dynamics and ecology, a common assumption is that individual fitness is higher at low densities because of lower intraspecific competition \cite{gotelli}. It has been realised, however, that populations can also display a positive correlation between its rate of growth and density, i.\,e., that at low population densities reproduction and survival of individuals may be smaller than at higher densities. This so-called Allee effect, first discussed in the 1930s \cite{allee}, can result from mate limitation, debilitated cooperative defense and feeding, unsubstantial predator satiation, dispersal, and habitat alteration, amongst other factors \cite{stephens,gascoigne,kramer,boukal,hastings}. Several species seem to display one form or another of Allee effect, such as the African wild dog (\textit{Lycaon pictus}), cod (\textit{Gadus morhua}) and marsh gentian (\textit{Gentiana pneumonanthe}); the effect, though, disappears as populations grow larger and does not affect most taxa.

\begin{figure}[t]
\centering
\includegraphics[viewport=90 65 460 410, scale=0.50, clip]{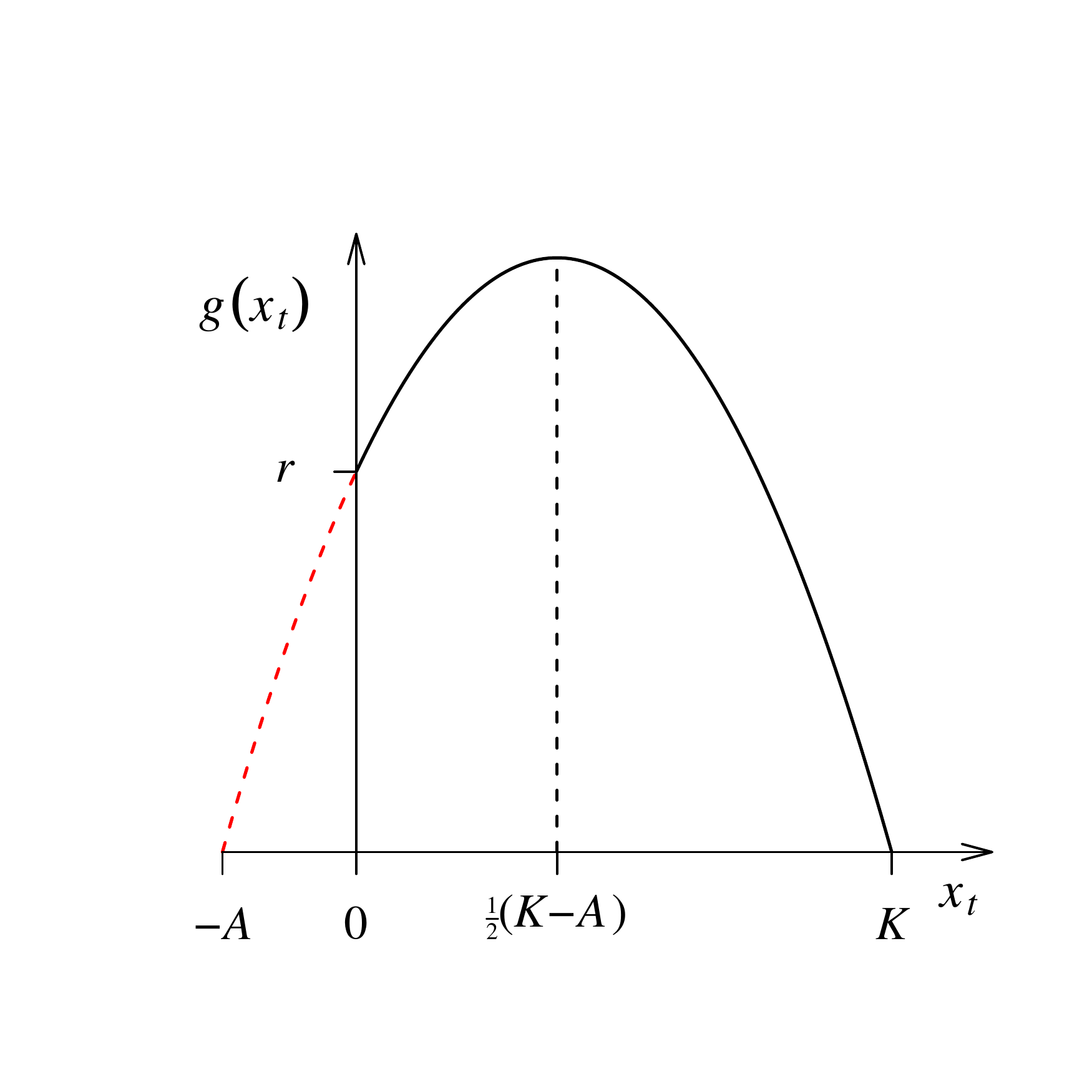}
\caption{\label{fig:weakallee}Example of intrinsic growth rate function $g(x_{t}) \geq 0$ displaying the combined features of logistic limitation to growth at high densities and weak Allee effect at low densities. Function $g(x_{t})$ is defined only for $x_{t} \geq 0$; the dashed red line does not have any meaning and is depicted only to locate $A$. See Sec.~\ref{sec:mfield} for the definition of the constants $r$, $K$ and $A$.}
\end{figure}

Allee effects can be weak (non-critical) or strong (critical). In the weak type, the intrinsic growth rate of the population decreases at small population densities but remains positive, whilst in the strong type the intrinsic growth rate may become negative and steer the population towards extinction. In discrete-time, the population density $x_{t}$ evolves through a map like $x_{t+1}=x_{t}g(x_{t})$ (see Sec.~\ref{sec:mfield}), such that the intrinsic growth rate function $g(x_{t})$ must be nonnegative and it only makes sense to speak of the weak Allee effect; see Figure~\ref{fig:weakallee}. In continuous-time, otherwise, $\dot{x}_{t}=x_{t}g(x_{t})$ and both types of Allee effect  can be modelled.

Let $p \in [0,1]$ be a real parameter describing the mean individual fitness (the ``survival strength'') in the population. Larger values of $p$ model more resilient individuals and favour reproduction. Our one-parameter PCA model for the dynamics of a single-species population under the influence of both logistic and weak Allee effects in terms of $p$ is given in Table~\ref{tab:p254q72}, where individuals are identified with $1$'s. The microscopic transition probabilities given in Table~\ref{tab:p254q72} embody the following rationale:
\begin{itemize} \setlength{\itemsep}{0ex}
\item[(i)]The transition $000 \to 010$ is forbidden on the basis that spontaneous generation of individuals is unnatural;
\item[(ii)]Logistic limitation to growth: birth rates $0 \to 1$ and survival $1 \to 1$ decrease with increased local density of individuals. The transition probabilities $\phi(1\,|\,101)=p$ and $\phi(1\,|\,111)=p$ reflect this limitation---individuals have to be fit to prosper in such crowded neighbourhoods;
\item[(iii)]Demographic Allee effect: birth rates $0 \to 1$ and survival $1 \to 1$ are hampered by low density of individuals. The transition probabilities $\phi(1\,|\,100) = \phi(1\,|\,010) = \phi(1\,|\,001) = p$ realise this effect;
\item[(iv)]Individuals that are neither lone nor in too packed a neighbourhood can endure indefinitely. This is modelled by the transition probabilities $\phi(1\,|\,110)=\phi(1\,|\,011)=1$.
\end{itemize}
Note that in item~(ii) we could have set $\phi(1\,|\,111)=0$ as well and that in item~(iv) we do not distinguish between an individual and its offspring as long as there is not an interruption (a death event) in its lifetime, making the model possibly useful also in the study of lineages.

The PCA in Table~\ref{tab:p254q72} is an example of a mixed PCA, also known as `diploid' cellular automata \cite{p182q200,asynchro,pxor,pcamap,Fates2017,DeBaets2018}. In a mixed PCA, two or more deterministic CA rules are combined such that sometimes one rule is applied, some other times another rule is applied. Asynchronous (or diluted) PCA are mixed PCA with one of the rules given by the identity map $x_{i}^{t+1}=x_{i}^{t}$. Reading the lines of Table~\ref{tab:p254q72} as binary numbers we see that our PCA is the mixed PCA $p254$--$q72$, with $q=1-p$.

\begin{table}[h]
\caption{\label{tab:p254q72}Rule table for PCA $p254$--$q72$ with $q=1-p$. The first row lists the initial neighbourhoods and the other two rows give the state reached in the next time step by the central cell with the probability given in the first column.}
\centering
\begin{tabular}{ccccccccc}
\hline\hline
 ${}$ & $111$ & $110$ & $101$ & $100$ & $011$ & $010$ & $001$ & $000$ \\ \hline
 $p$  &  $1$  &  $1$  &  $1$  &  $1$  &  $1$  &  $1$  &  $1$  &  $0$  \\
$1-p$ &  $0$  &  $1$  &  $0$  &  $0$  &  $1$  &  $0$  &  $0$  &  $0$  \\
\hline\hline
\end{tabular}
\end{table}


\section{\label{sec:mfield}Single cell mean field approximation}

From (\ref{eq:p1}), (\ref{eq:mf}), and Table~\ref{tab:p254q72}, the single-cell mean field equation for the time evolution of $P_{t}(x=1) \equiv x_{t}$ (and so $P_{t}(x=0) \equiv 1-x_{t}$) of PCA $p254$--$q72$ reads
\begin{equation}
\label{eq:xxx}
x_{t+1} = px_{t}^{3}+(2+p)x_{t}^{2}(1-x_{t})+3px_{t}(1-x_{t})^{2}.
\end{equation}
Equation (\ref{eq:xxx}) has the expected structure for discrete-time models of population dynamics, to wit, $x_{t+1}=x_{t}g(x_{t})$, with $g(x_{t}) \geq 0$ the intrinsic growth rate function of the population. Every PCA with $\phi(1\,|\,000)=0$, i.\,e., with $\bm{x}=(0,0,\cdots,0)$ an absorbing state, will display the same structure. At $p=2/3$, (\ref{eq:xxx}) becomes the logistic map
\begin{equation}
\label{eq:logis}
x_{t+1}=rx_{t}\Big(1-\frac{x_{t}}{K}\Big)
\end{equation}
with $r=2$ representing the maximum potential rate of reproduction of the individuals in the population and $K=3/2$ the carrying capacity, i.\,e., the maximum population viable under the given ecological conditions \cite{gotelli}. The logistic map (\ref{eq:logis}) quickly converges (for $0 < x_{0} \leq 1$) to the stable population $\lim_{t \to \infty} x_{t}=3/4$.

\begin{figure}[t]
\centering
\includegraphics[viewport=0 0 540 430, scale=0.55, clip]{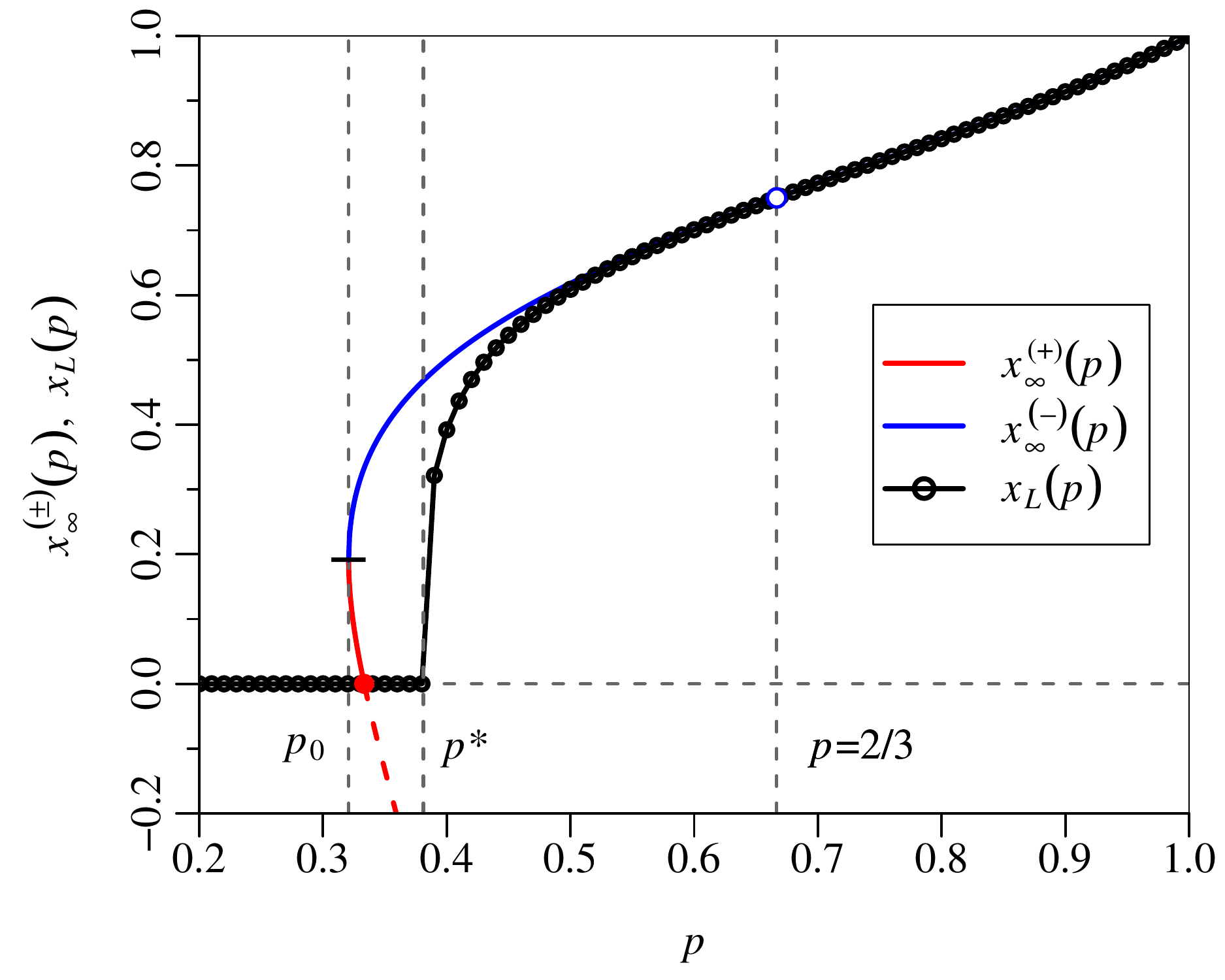}
\caption{\label{fig:mfield}Critical points and density profiles $\mathrm{Re}\{x_{\infty}^{(\pm)}(p)\}$ from the mean field solution (\ref{eq:xpm}) and $x_{L}(p)$ from direct Monte Carlo simulations of a PCA with $L=10\,000$ cells averaged over $10\,000$ samples. The mean field solution predicts a first-order phase transition at the critical point is $p_{\mathrm{mf}}^{*}=p_{0}=$ {$(8-2\sqrt{5})/{11} \simeq 0.321$}, whilst the empirical critical point $p^{*} \simeq 0.381$ (see Sec.~\ref{sec:critical}). The mean field solution $x_{\infty}^{(-)}(p)$ and the empirical profile $x_{L}(p)$ are virtually equal for $p \gtrsim 0.5$.}
\end{figure}

Back to our cubic map (\ref{eq:xxx}), in the stationary state we must have $x_{t+1}=x_{t} \equiv x_{\infty}$ and (\ref{eq:xxx}) has solutions $x_{\infty}^{(0)}=0$, corresponding to the absorbing state devoid of individuals, and
\begin{equation}
\label{eq:xpm}
x_{\infty}^{(\pm)}=\frac{(5p-2) \pm \sqrt{(5p-2)^{2}-4(3p-2)(3p-1)}}{2(3p-2)}.
\end{equation}
Solutions (\ref{eq:xpm}) are real for $p \geq p_{0} = (8-2\sqrt{5})/{11} \simeq 0.321$, at which $x_{\infty}^{(\pm)}(p_{0})=(3-\sqrt{5})/4 \simeq 0.191$. Solution $x_{\infty}^{(+)}$ is positive between this point and its root at $p=\frac{1}{3}$, diverges and changes sign at $p=\frac{2}{3}$, and never becomes $x_{\infty}^{(+)} \leq 1$ again. Solution $x_{\infty}^{(-)}$ does not have any real root, is positive for all $p \geq p_{0}$ with a minimum at $p_{0}$ and with $x_{\infty}^{(-)}(\frac{1}{3})=\frac{1}{3}$. The point $p=\frac{2}{3}$ is a removable discontinuity of $x_{\infty}^{(-)}$, at which $x_{\infty}^{(-)}(\frac{2}{3})=\frac{3}{4}$. The mean field solutions (\ref{eq:xpm}) are summarised in Figure~\ref{fig:mfield}. The single-cell mean field approximation thus predicts a first-order phase transition for PCA $p254$--$q72$ at $p_{\mathrm{mf}}^{*} = p_{0} = (8-2\sqrt{5})/{11} \simeq 0.321$.

\begin{figure}[t]
\centering
\includegraphics[viewport=90 60 470 450, scale=0.62, clip]{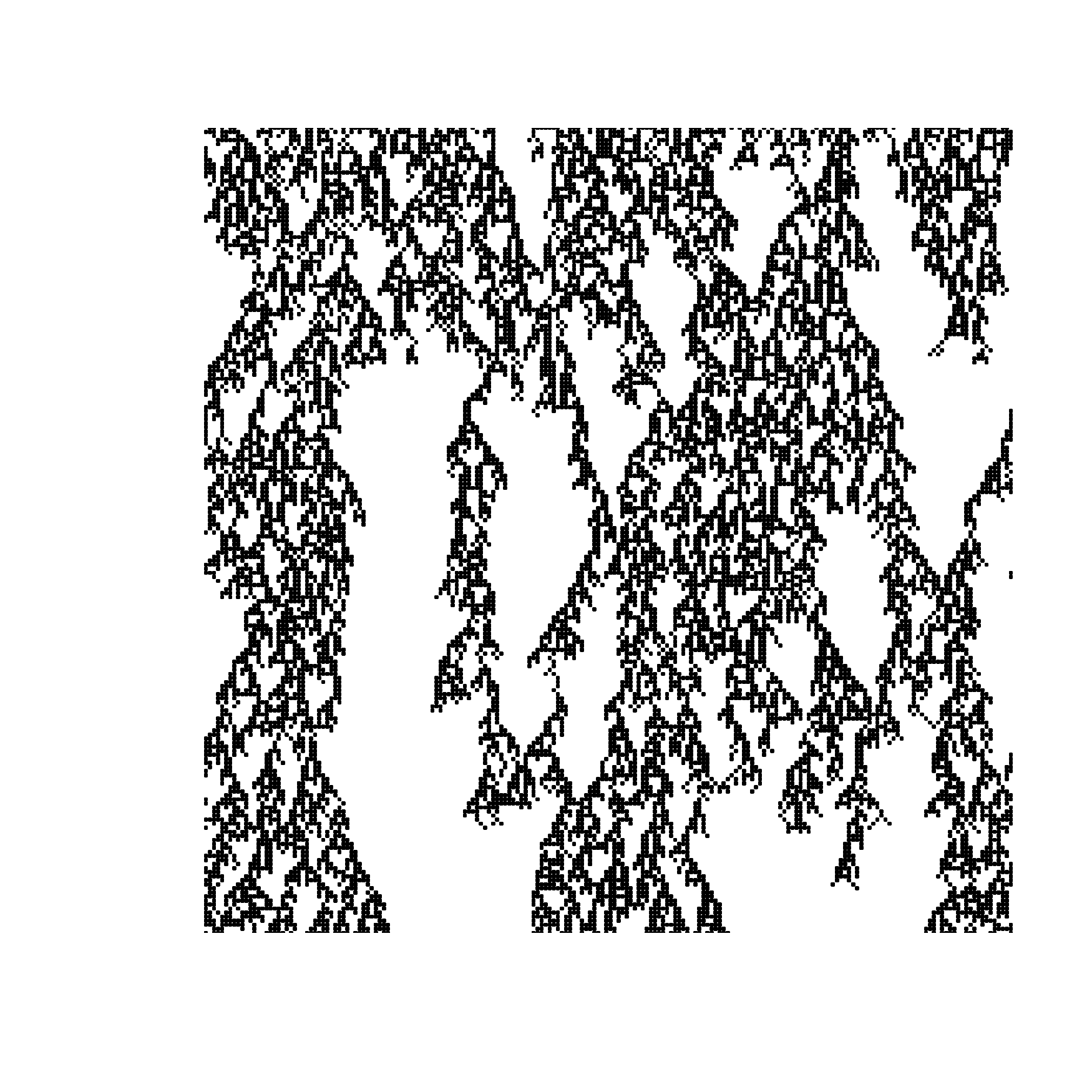}
\caption{\label{fig:diagram}Space-time diagram of PCA $p254$--$q72$ at $p=7/18 \simeq 0.389 > p^{*} \simeq 0.381$ (see Sec.~\ref{sec:critical}). A total of $L=200$ cells under periodic boundary conditions are evolved for $200$ time steps (from top to bottom) from an initially random state of density $\sim 2/3$.}
\end{figure}

The mean field maps (\ref{eq:xxx}) and (\ref{eq:logis}) can be compared with a discrete-time model for the dynamics of populations under weak Allee effects that reads \cite{boukal,hastings,pcamap}
\begin{equation}
\label{eq:allee}
x_{t+1} = rx_{t}\Big(1-\frac{x_{t}}{K}\Big)\Big(1+\frac{x_{t}}{A}\Big) =
x_{t}g(x_{t}),
\end{equation}
with $0<A<K$ is a critical population threshold parameter. The instrinsic growth rate function $g(x_{t})$ corresponding to (\ref{eq:allee}) is displayed in Figure~\ref{fig:weakallee}. After simple algebra, we see that map (\ref{eq:xxx}) can be recast like (\ref{eq:allee}) as long as $0 \leq p < \frac{2}{5}$, with
\begin{equation}
\label{eq:rKA}
r=3p, \quad K-A=(2-5p)/(2-3p), \quad KA=3p/(2-3p).
\end{equation}
For example, if we set $p=7/18 > p^{*} > p_{\mathrm{mf}}^{*}$ (we will find in Sec.~\ref{sec:critical} that the empirical critical point $p^{*} \simeq 0.381$), but still within $[0,\frac{2}{5})$, we obtain $r=7/6$, $K=(\sqrt{1261}+1)/30 \simeq 1.217$ and $A=(\sqrt{1261}-1)/30 \simeq 1.150$. Figure~\ref{fig:diagram} displays the stationary pattern of a PCA $p254$--$q72$ with $L=200$ cells at $p=7/18$. The average stationary density measured from simulations ($L=10\,000$, average taken over $10\,000$ samples) is $\langle x_{L}(7/18) \rangle \simeq 0.310$, whilst the stationary density obtained from (\ref{eq:xpm}) (or (\ref{eq:allee})--(\ref{eq:rKA})) is $x_{\infty}^{(-)}(7/18) = (\sqrt{181}+1)/30 \simeq 0.482$. The mean field solution clearly overestimates the stationary density near the critical point $p^{*}$, where fluctuations are large. Note that when $p \to 0$ in (\ref{eq:rKA}), both $r \to 0$ and $A \to 0$ such that $r/A \to 2$ and $K \to 1$. Indeed, when $A \to 0$ we have $1+x_{t}/A \approx x_{t}/A$ and the right-hand side of (\ref{eq:allee}) becomes $(r/A)x_{t}^{2}(1-x_{t}/{K})$, which models a particular case of weak Allee effect.

\section{\label{sec:critical}Critical behaviour}

\subsection{Critical indices}

We determine the critical behaviour of PCA $p254$--$q72$ by Monte Carlo simulations and finite-size scaling analysis of the data. The techniques are standard and can be reviewed in \cite{DickMarro99,Haye00}. The main quantity of interest is the empirical time-dependent density of active cells
\begin{equation}
x_{L}(t)=\frac{1}{L}\sum_{i=1}^{L}x_{i}^{t}.
\end{equation}
Close to the critical point $p=p^{*}$, we expect that
\begin{equation}
\label{eq:scaling}
x_{L}(t) \sim t^{-\beta/\nu_{\|}}\,\Phi(\epsilon t^{1/\nu_{\|}},\,t^{\nu_{\perp}/\nu_{\|}}/L),
\end{equation}
where $\epsilon=\abs{p-p^{*}} \geq 0$ and $L$ is the size of the array. The exponents $\nu_{\|}$ and $\nu_{\perp}$ rule the scaling behaviour of the PCA at criticality in the temporal ($\nu_{\|}$) and spatial ($\nu_{\perp}$) dimensions. For large systems ($L \nearrow \infty$), $x_{L}(t) \sim t^{-\beta/\nu_{\|}}\Phi(\epsilon t^{1/\nu_{\|}})$, with $\Phi(u \ll 1) \sim \mathrm{const}$ and $\Phi(u \gg 1) \sim u^{\beta}$. Close to the critical point $\epsilon \approx 0$ and for large $L$ we must thus observe $x_{L}(t) \sim t^{-\delta}$, with $\delta=\beta/\nu_{\|}$, and one can simultaneously determine $p^{*}$ and $\delta$ by inspection of logarithmic plots of $x_{L}(t)$.

\begin{figure}[t]
\centering
\includegraphics[viewport=5 10 538 430, scale=0.42, clip]{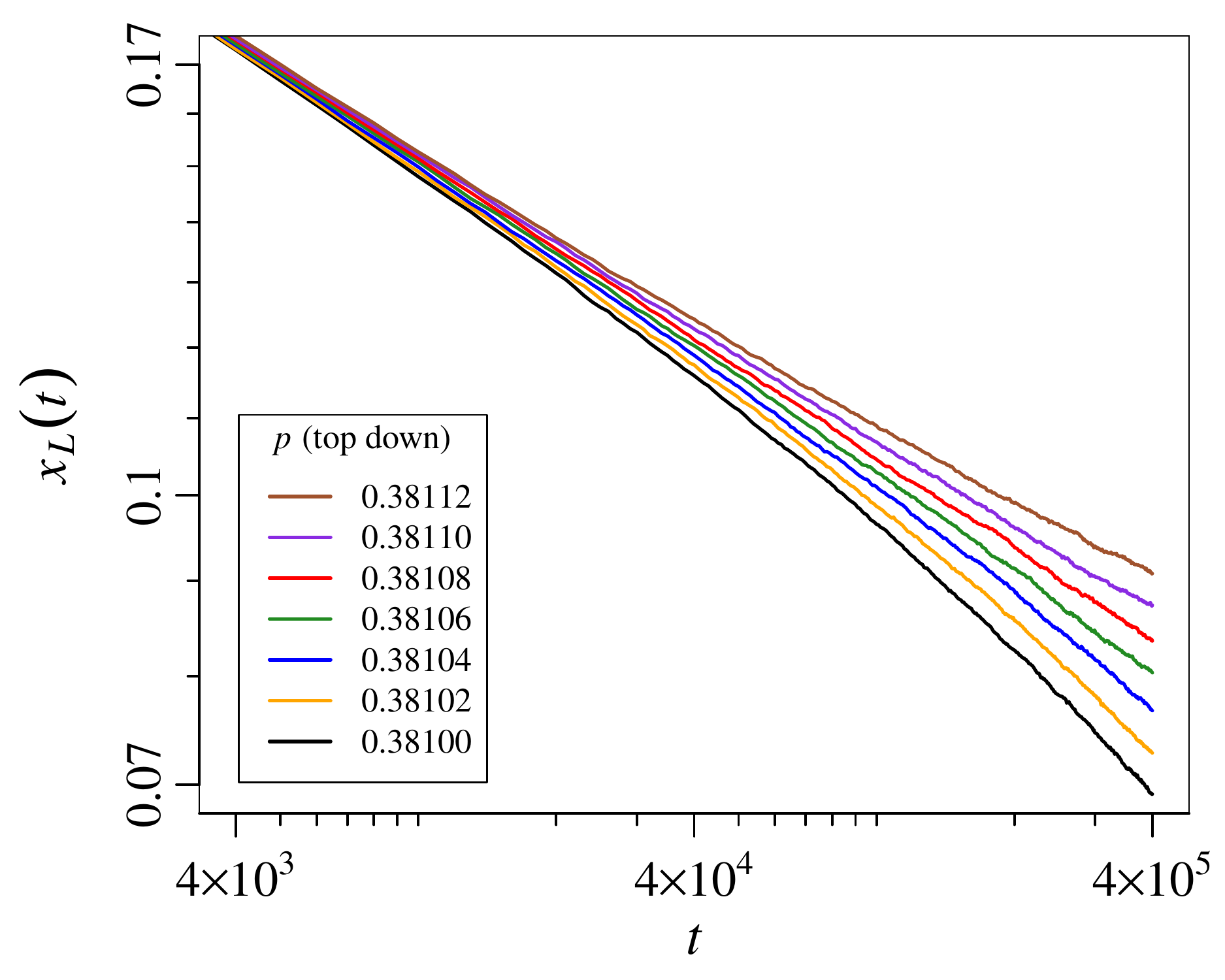} \hfill
\includegraphics[viewport=5 10 538 430, scale=0.42, clip]{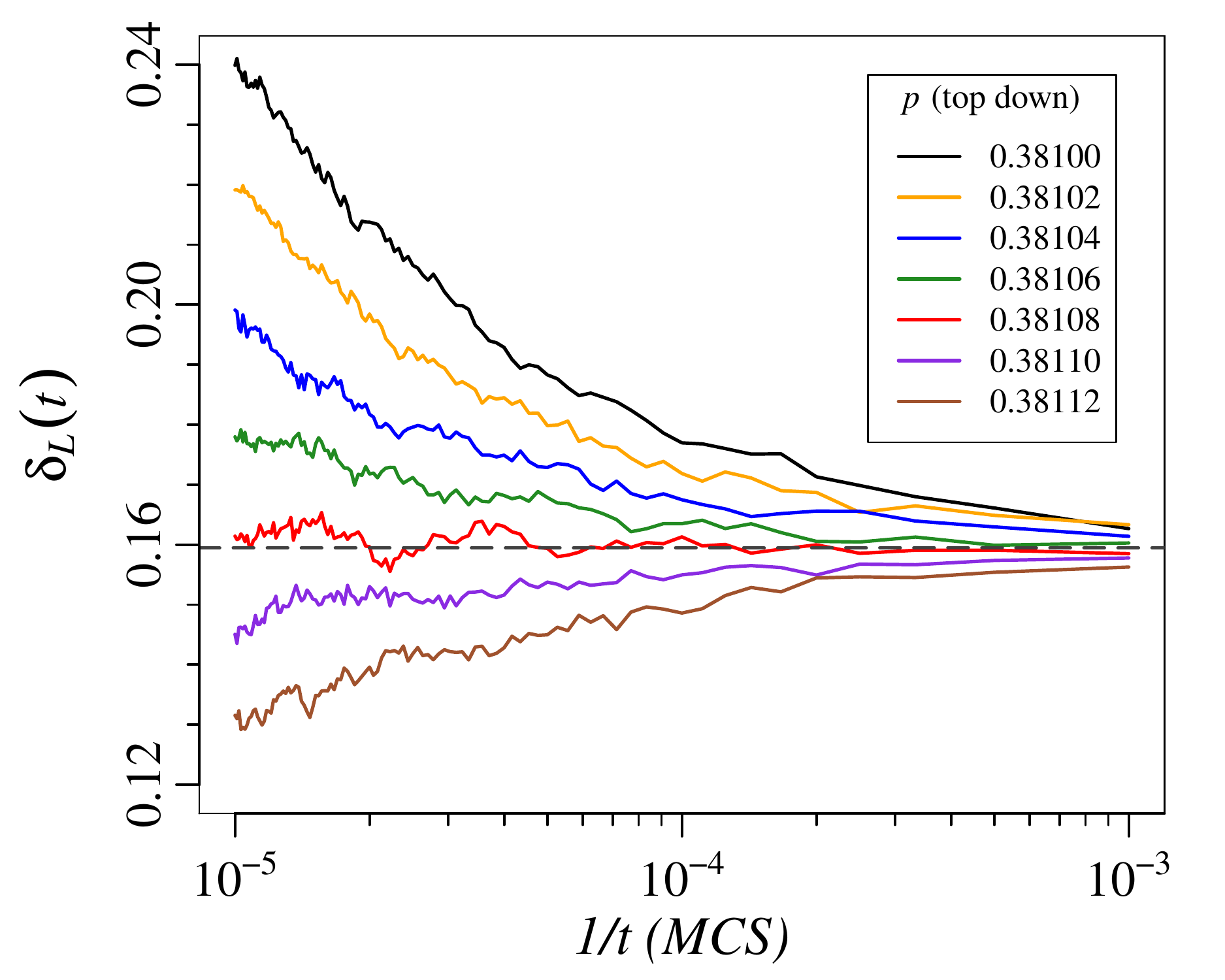}
\caption{\label{fig:delta}Left panel: Logarithmic plot of $x_{L}(t)$ for a PCA with $L=20\,000$ cells; each curve is an average over $10\,000$ samples. Right panel: effective exponent $\delta_{L}(t)$ obtained from the curves in the left panel. From these curves we estimate $p^{*}=0.38108(1)$ and $\delta=0.161(2)$. The dashed line in the right panel indicates the best known value for the corresponding exponent of the directed percolation (DP) process on the square lattice, $\delta_{\mathrm{DP}}=0.159\,464(6)$.}
\end{figure}

Figure~\ref{fig:delta} displays $x_{L}(t)$ for a PCA of $L=20\,000$ cells and $4000 \leq t \leq 400\,000$; each curve is an average over $10\,000$ samples. We see that for $p<p^{*}$ the curves bend downward, indicating that the population heads to extinction, whilst for $p>p^{*}$ the curves bend upward and the population will eventually reach a stable density. At the critical point the decay must be scale-invariant, i.\,e., algebraic. The right panel displays the effective exponent
\begin{equation}
\label{eq:delta}
\delta_{L}(t)=\log_{\,b}[x_{L}(t)/x_{L}(bt)]
\end{equation}
against $1/t$. We take $b=4$ in our analyses. From Figure~\ref{fig:delta} we estimate $p^{*}=0.38108(1)$ and $\delta=0.161(2)$, where the numbers between parentheses indicate the uncertainty in the last digit of the data. The estimate for $\delta$ comes from an extrapolation of the form $\delta_{L}(t) = \delta_{L}+at^{-1}$ to the data for $L=20\,000$ and $p=0.38108$, with a conservative estimate for the uncertainty.

\begin{figure}[t]
\centering
\includegraphics[viewport=5 10 538 430, scale=0.42, clip]{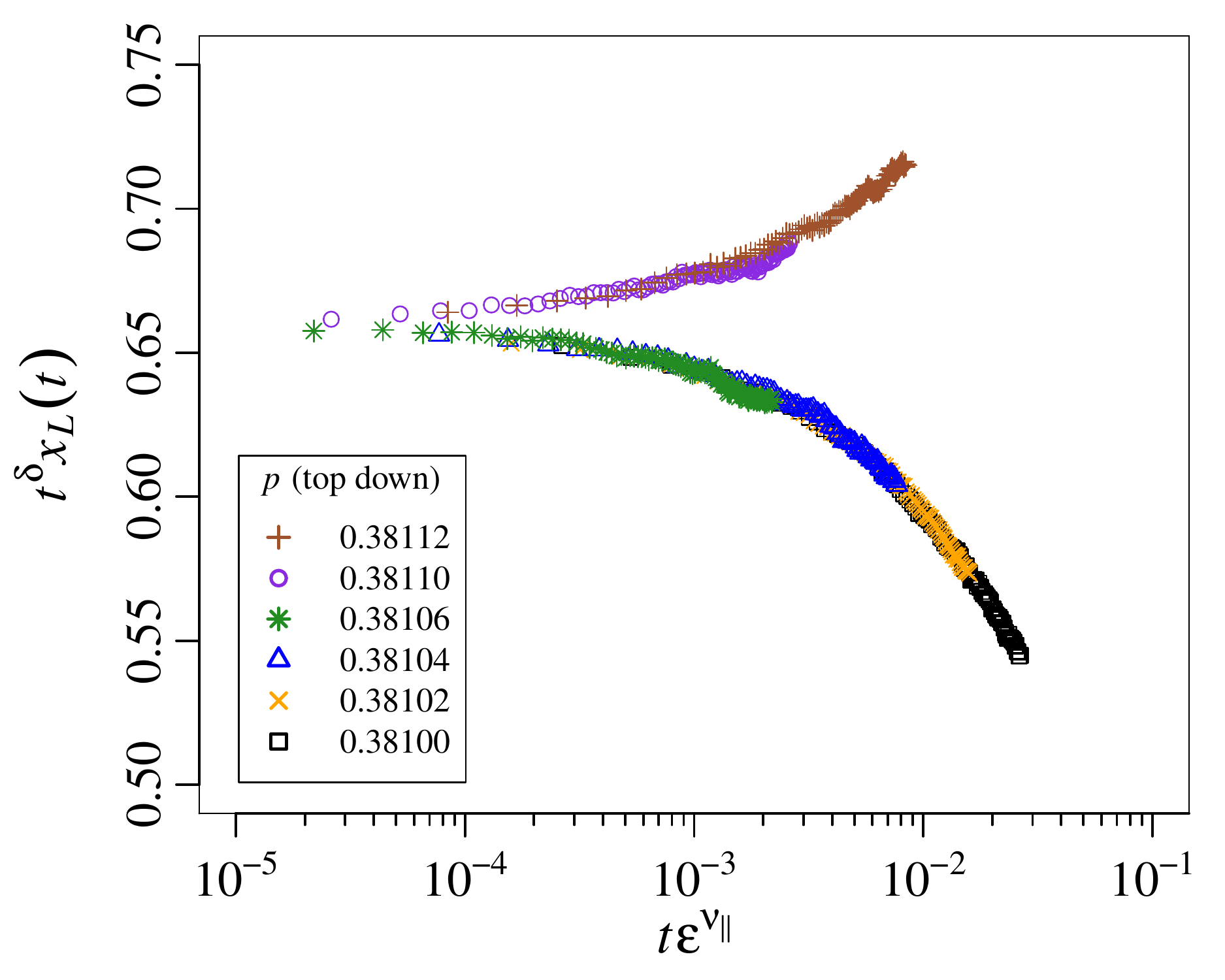} \hfill
\includegraphics[viewport=5 10 538 430, scale=0.42, clip]{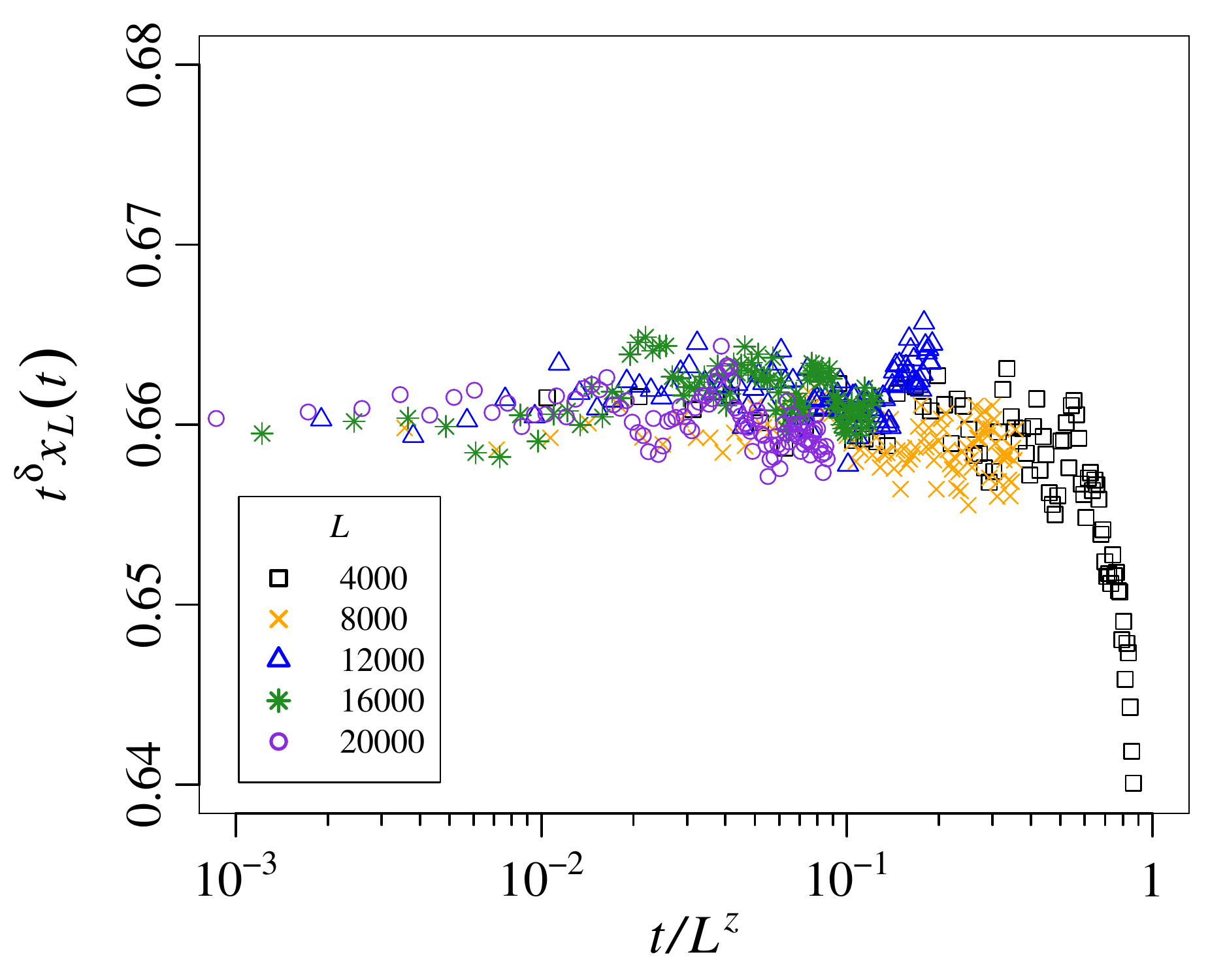}
\caption{\label{fig:zeta}Data collapse for the off-critical (left panel) and finite-size (right panel) data, obtained with $\nu_{\|}=1.75$, $z=1.55$ and $\delta=0.160$, where $\epsilon=\abs{p-p^{*}}$ with $p^{*}=0.38108$. In the left panel, $L=20\,000$ and the upper (lower) branch corresponds to $p>p^{*}$ ($p<p^{*}$). In the right panel, the late times data for $L=4000$ do not quite bunch together with the other data, although the data for smaller $t$ do. Note that whilst the data is spread over $\sim 3$ decades in the $x$-axis, the range of the $y$-axis is relatively tight.}
\end{figure}

The exponent $\nu_{\|}$ can be obtained by plotting $t^{\delta}x_{L}(t)$ against $t\epsilon^{\nu_{\|}}$ and tuning $\nu_{\|}$ to achieve data collapse with different $\epsilon$---a so-called off-critical estimation. In the same way, by plotting $t^{\delta}x_{L}(t)$ versus $t/L^{z}$ at criticality for different $L$ and tuning $z$ until data collapse provides an estimate for the exponent $z$---a finite-size scaling estimation. Since $z=\nu_{\|}/\nu_{\perp}$, the procedure also furnishes an estimate for $\nu_{\perp}$. We achieved best data collapse with $\nu_{\|}=1.75$, $z=1.55$ and $\delta=0.160$. We could not discern the values of these exponents more precisely than by $\pm 0.05$, $\pm 0.05$ and $\pm 0.001$, respectively. The data collapse is, otherwise, very sensitive to $p^{*}$. The finite-size curves using the above stated values for $\nu_{\|}$, $z$ and $\delta$ (and $p^{*} = 0.38108$) appear in Figure~\ref{fig:zeta}.

The three exponents, $\delta$, $\nu_{\|}$, and $z$, suffice to determine the universality class of critical behaviour of the model, the other exponents
following from known hyperscaling relations \cite{DickMarro99,Haye00}. The best values available for the critical exponents of the directed percolation (DP) process on the square lattice (or, equivalently, the ${(1+1)}$-dimensional basic contact process) are $\delta_{\mathrm{DP}}=0.159\,464(6)$, $\nu_{\|\mathrm{DP}}=1.733\,847(6)$, and $z_{\mathrm{DP}}=1.580\,745(10)$ \cite{jensen,munoz,precise}. Thus, within error bars our data put the critical behaviour of PCA $p254$--$q72$ in the DP universality class.

\subsection{Remarks on the directed percolation transition}

Directed percolation entered the lore of population dynamics and mathematical ecology mostly through the study of the dynamics of populations about the threshold of ecological extinction \cite{Bascompte96,Boswell98,Szabo05,%
Szabo07,Davis08}. The spatial dynamics of the population, driven by locally density-dependent rates of dispersal and reproduction, may lead (even in an homogeneous, contiguous environment) to isolated clusters of individuals that may thrive or not, eventually leading either to the overall survival or extinction of the population.

A paradigmatic model with critical behaviour in the DP universality class is the basic contact process (CP), see \cite{DickMarro99} and the references therein. The basic CP is an interacting particle system on the lattice evolving in continuous-time by the transitions $1 \to 0$ and $01$ or $10 \to 11$ with reaction rates $1$ and $\lambda$, respectively, with the later involving only nearest neighbours on the lattice. In one spatial dimension, the basic CP displays an extinction-survival phase transition at the critical point $\lambda^{*} \simeq 1.649$ \cite{jensen,munoz,precise}. The first order mean field equation for the dynamics of the density of individuals is given by the logistic equation $\dot{x}_{t}=2{\lambda}x_{t}(1-x_{t})-x_{t}$, with a stationary density given either by $x_{\infty}=0$ or $x_{\infty}=1-\frac{1}{2}\lambda^{-1}$, $\lambda \geq \frac{1}{2}$, with a continuous transition between the two phases taking place at the mean field critical point $\lambda^{*}_{\mathrm{mf}}=\frac{1}{2}$. Clearly, the basic CP can be applied to many models of epidemiology and population dynamics \cite{Levin94,Levin96,Pacala97}, including generalisations to multiple-species predator-prey models \cite{Mobilia2007,Uwe2018,Tauber14}. DP critical behaviour has also been found in the extinction of certain types of strategies in spatial evolutionary games \cite{SzToke98,Hauert05,Vukov08}. From the point of view of theoretical physics, that the critical behaviour in these models belongs to the DP universality class may be viewed as an expected consequence of the so-called DP conjecture, according to which phase transitions into an absorbing state in short-ranged systems in the absence of conserved quantities belong to the DP universality class of critical behaviour \cite{janssen,grassberger}. Critical behaviour in the DP universality class appears in many stylised mathematical models of natural phenomena, despite its lack of experimental evidence until recently \cite{Takeuchi09,Lemoult16}.

PCA $p254$--$q72$ can be viewed as yet another model displaying DP critical behaviour. Its distinguishing feature, however, is that its first order mean field approximation (\ref{eq:xxx}) to the dynamics of the density of individuals is a cubic map, not a quadratic one like the logistic map, with the cubic term representing a weak Allee factor. It is remarkable that the rationale presented in Section~\ref{sec:mixed} for the construction a model for a population of individuals simultaneously under the stresses of overcrowding and loneliness recovers a logistic-like map including the weak Allee effect. Although the mean field equation for the dynamics of a one-dimensional PCA of symmetric radius $r=1$ is almost always a cubic map and the analysis of cubic maps from PCA can be found in the literature, (i)~not every choice of transition probabilities ensue a cubic map with the correct sign, as detailed in \cite{pcamap}, and (ii)~the connexion between mixed PCA, cubic maps and Allee effects seems to be novel and relevant.


\section{\label{sec:summary}Summary and conclusions}

Recently, all six-parameter left-right symmetric elementary (two-state) PCA that recover the logistic map with or without weak Allee effect from their first-order mean field approximation have been identified \cite{pcamap}, including their one-parameter counterparts. The one-parameter PCA found, however, have a particular strucure: they are all mixed (or `diploid') PCA $pA$--$qB$ with $A$ and $B$ even and $A+B=254$. That means that the bits $A_{i}$ and $B_{i}$ of the rules $A$ and $B$ are such that $A_{0}=B_{0}=0$, because $\phi(1\,|\,000)=0$, and $B_{i}=1-A_{i}$, $i=1, \dots, 7$. PCA $p254$--$q72$ does not fit into such PCA, because the transition probabilities $\phi(1\,|\,011)=\phi(1\,|\,110)=1$ (see Table~\ref{tab:p254q72}), making bits $A_{3}=B_{3}=1$ and $A_{6}=B_{6}=1$ simultaneously; it thus belongs to another set of mixed PCA that yields logistic-like maps like (\ref{eq:allee}) including the weak Allee effect in the first-order mean field approximation \textit{and} displays an extinction-survival phase transition. Preliminary results indicate that mixed PCA displaying extinction-survival phase transitions abound, and some effort is currently being made to spot patterns in the mess \cite{pcamap,Fates2017,DeBaets2018}. Of particular interest to us is the subclass of mixed PCA that yields cubic maps like (\ref{eq:allee}) in first-order mean field approximation, together with the analysis of the maps themselves. 

\begin{figure}[t]
\centering
\includegraphics[viewport=5 10 538 430, scale=0.42, clip]{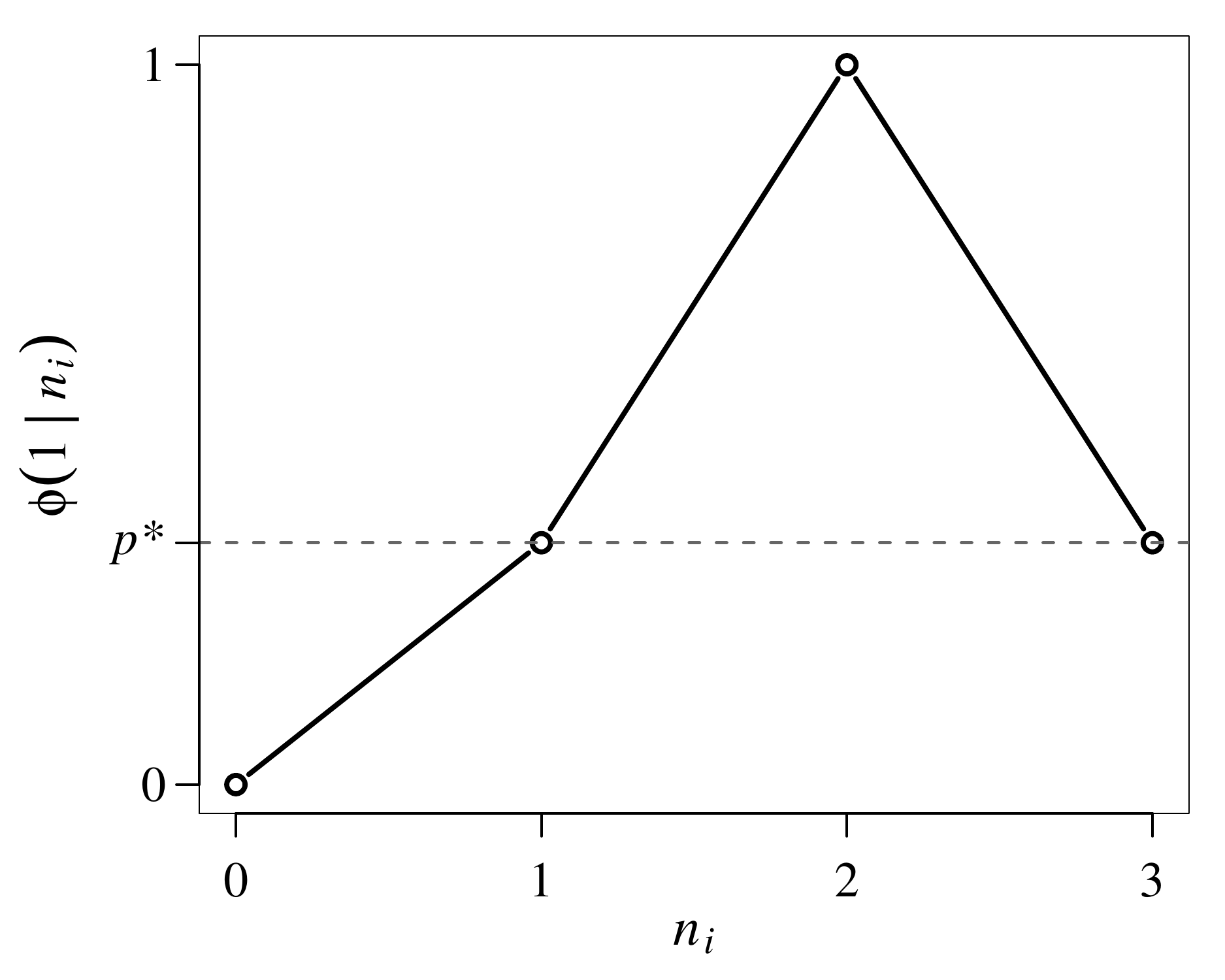}
\caption{\label{fig:totalistic}Microscopic transition probabilities $\phi(1\,|\,n_{i})$ for the semi-totalistic PCA $p254$--$q104$. The transition probabilities mimic locally the combined logistic and weak Allee effects. PCA $p254$--$q104$ displays an extinction-survival phase transition at $p^{*} \simeq 0.336$ in the DP universality class of critical behaviour.}
\end{figure}

Let us mention an \textit{apropos} semi-totalistic PCA embodying the rationale given in Section~\ref{sec:mixed}. In a semi-totalistic PCA, the transition probabilities depend only on the total number $n_{i}$ of occupied cells in the neighbourhood of cell $x_{i}$, including $x_{i}$ itself; in our case, $n_{i}=x_{i-1}+x_{i}+x_{i+1}$. If we take the transition probabilities $\phi(1\,|\,n_{i})$ to mimic locally the combined logistic and weak Allee effects according to Figure~\ref{fig:totalistic}, we end up with the mixed PCA $p254$--$q104$ of Table~\ref{tab:p254q104}, which displays an extinction-survival phase transition at $p^{*} \simeq 0.336$ in the DP universality class of critical behaviour. If we alternatively set $\phi(1\,|\,3)=0$, we get PCA $p126$--$q104$ with a critical point at $p^{*} \simeq 0.386$, also in the DP universality class of critical behaviour.

PCA $p254$--$q72$ and its siblings (PCA $p254$--$q104$, $p126$--$q104$, and $p126$--$q72$, this last one with a $p^{*} \simeq 0.416$) model the logistic limitation to growth and weak Allee effect by means of a single parameter. The PCA would be more versatile if we could model these two effects by two independent parameters, say, $\lambda$ and $\alpha$ denoting respectively the intensities of the logistic limitation and weak Allee effects. Then, for example, one could take $\phi(1\,|\,101)=1-\lambda$ but $\phi(1\,|\,010)=1-\alpha$. A two-dimensional version of the PCA and its possible relationship with invasion dynamics would also be of interest. We can reasonably guess that there will be an extinction-survival phase transition in this case, but the characteristics of the transition, the scaling of the boundaries, and their biological relevance and interpretation remains to be investigated.


\begin{table}
\caption{\label{tab:p254q104}Rule table for the semi-totalistic PCA $p254$--$q104$; the table reads like Table~\ref{tab:p254q72}. The transition probabilities $\phi(1\,|\,n_{i})$ are given by $\phi(1\,|\,0)=0$, $\phi(1\,|\,1)=p$, $\phi(1\,|\,2)=1$, and $\phi(1\,|\,3)=p$, where $n_{i}=x_{i-1}+x_{i}+x_{i+1}$.}
\centering
\begin{tabular}{ccccccccc}
\hline\hline
 ${}$ & $111$ & $110$ & $101$ & $100$ & $011$ & $010$ & $001$ & $000$ \\ \hline
 $p$  &  $1$  &  $1$  &  $1$  &  $1$  &  $1$  &  $1$  &  $1$  &  $0$  \\
$1-p$ &  $0$  &  $1$  &  $1$  &  $0$  &  $1$  &  $0$  &  $0$  &  $0$  \\
\hline\hline
\end{tabular}
\end{table}


\section*{Acknowledgments}

We thank M\'{a}rio J de Oliveira (USP) and Yeva Gevorgyan (USP) for useful conversations and three anonymous reviewers for constructive comments improving the manuscript. This work was partially supported by the S\~{a}o Paulo State Research Foundation -- FAPESP (Brazil) through research grant 2015/21580-0.



\vspace{4ex}

\centerline{$\bm{*}$ --- $\bm{*}$ --- $\bm{*}$}


\begin{thebibliography}{99} 

\bibitem{JvN1966}von Neumann J 1966 \textit{Theory of Self-Reproducing Automata} ed W A Burks (Urbana: University of Illinois Press)

\bibitem{Wolfram94}Wolfram S 1994 \textit{Cellular Automata and Complexity: Collected Papers} (Reading, MA: Addison-Wesley)

\bibitem{Boccara10}Boccara N 2010 \textit{Modeling Complex Systems} 2nd ed (New York: Springer)

\bibitem{Adamatzky13}Adamatzky A 2013 \textit{Reaction-Diffusion Automata: Phenomenology, Localisations, Computation} (Berlin: Springer)

\bibitem{Russkiye90}Toom A L, Vasilyev N B, Stavskaya O N, Mityushin L G, Kurdyumov G L and Pirogov S A 1990 Discrete local Markov systems \textit{Stochastic Cellular Systems: Ergodicity, Memory, Morphogenesis} ed R~L~Dobrushin, V~I~Kryukov and A~L~Toom (Manchester: Manchester University Press) pp~1--182



\bibitem{Bennett85}Bennett C H and Grinstein G 1985 Role of irreversibility in stabilizing complex and nonergodic behavior in locally interacting discrete systems \textit{Phys. Rev. Lett.} \textbf{55} 657--60

\bibitem{Gacs01}G\'{a}cs P 2001 Reliable cellular automata with self-organization \textit{J. Stat. Phys.} \textbf{103} 45--267

\bibitem{Hogeweg88}Hogeweg P 1988 Cellular automata as a paradigm for ecological modeling \textit{Appl. Math. Comput.} \textbf{27} 81--100

\bibitem{Caswell93}Caswell H and Etter R J 1993 {Ecological interactions in patchy environments: From patch-occupancy models to cellular automata} \textit{Patch Dynamics} ed S~A~Levin, T~M~Powell and J~W~Steele (Berlin: Springer) pp~93--109

\bibitem{Molofsky94}Molofsky J 1994 Population dynamics and pattern formation in theoretical populations \textit{Ecology} \textbf{75} 30--9

\bibitem{Durrett94}Durrett R and Levin S A 1994 Stochastic spatial models: A user's guide to ecological applications \textit{Phil. Trans. R. Soc. London} B \textbf{343} 329--50

\bibitem{RandWilson95}Rand D and Wilson H 1995 Using spatio-temporal chaos and intermediate-scale determinism to quantify spatially extended ecosystems \textit{Proc. R. Soc. Lond.} B \textbf{259} 111--7

\bibitem{Kareiva1997}Tilman D, Lehman C L and Kareiva P M 1997 Population dynamics in spatial habitats \textit{Spatial Ecology: The Role of Space in Population Dynamics and Interspecific Interactions} ed D~Tilman and P~M~Kareiva (Princeton: Princeton University Press) pp~3--30

\bibitem{Baltzer98}Balzter H, Braun P W and K\"{o}hler W 1998 Cellular automata models for vegetation dynamics \textit{Ecol. Model.} \textbf{107} 113--25

\bibitem{Silvertown92}Silvertown J, Holtier S, Johnson J and Dale P 1992 Cellular automaton models of interspecific competition for space--the effect of pattern on process \textit{J. Ecol.} \textbf{80} 527--33


\bibitem{Turner01}Turner M G, Gardner R H and O'Neill R V 2001 \textit{Landscape Ecology in Theory and Practice: Pattern and Process} (New York: Springer)

\bibitem{Wagner97}Wagner D 1997 Cellular automata and geographic information systems \textit{Environ. Plan.} B \textbf{24} 219--34

\bibitem{Benenson04}Benenson I and Torrens P M 2004 \textit{Geosimulation: Automata-based Modeling of Urban Phenomena} (West Sussex: Wiley)


\bibitem{gotelli}Gotelli N J 2008 \textit{A Primer of Ecology} 4th ed (Sunderland, MA: Sinauer)

\bibitem{allee}Allee W C 1931 \textit{Animal Aggregations: A Study in General Sociology} (Chicago: University of Chicago Press)


\bibitem{stephens}Stephens P A, Sutherland W J and Freckleton R P 1999 What is the Allee effect? \textit{Oikos} \textbf{87} 185--90


\bibitem{gascoigne}Courchamp F, Berec L and Gascoigne J 2008 \textit{Allee Effects in Ecology and Conservation} (Oxford: Oxford University Press)

\bibitem{kramer}Kramer A M, Dennis B, Liebhold A M and Drake J M 2009 The evidence for Allee effects \textit{Popul. Ecol.} \textbf{51} 341--54

\bibitem{boukal}Boukal D S and Berec L 2002 Single-species models of the Allee effect: Extinction boundaries, sex ratios and mate encounters \textit{J. Theor. Biol.} \textbf{218} 375--94

\bibitem{hastings}Taylor C M and Hastings A 2005 Allee effects in biological invasions \textit{Ecol. Lett.} \textbf{8} 895--908

\bibitem{local}Gutowitz H A, Victor J D and Knight B W 1987 Local structure theory for cellular automata \textit{Physica} D \textbf{28} 18--48

\bibitem{DickMarro99}Marro J and Dickman R 1999 \textit{Nonequilibrium Phase Transitions in Lattice Models} (Cambridge, UK: Cambridge University Press)

\bibitem{p182q200}Mendon\c{c}a J R G and de Oliveira M J 2011 An extinction-survival-type phase transition in the probabilistic cellular automaton $p182$--$q200$ \textit{J. Phys. A: Math. Theor.} \textbf{44} 155001

\bibitem{asynchro}Fat\`{e}s N 2014 A guided tour of asynchronous cellular automata \textit{J. Cell. Autom.} \textbf{9} 387--416

\bibitem{pxor}Mendon\c{c}a J R G 2016 The inactive-active phase transition in the noisy additive (exclusive-or) probabilistic cellular automaton \textit{Int. J. Mod. Phys.} C \textbf{27} 1650016

\bibitem{pcamap}Mendon\c{c}a J R G and Gevorgyan Y 2017 Approximate probabilistic cellular automata for the dynamics of single-species populations under discrete logisticlike growth with and without weak Allee effects \textit{Phys. Rev.} E \textbf{95} 052131


\bibitem{Fates2017}Fat\`{e}s N 2017 Diploid cellular automata: First experiments on the random mixtures of two elementary rules \textit{Cellular Automata and Discrete Complex Systems -- AUTOMATA~2017} ed A Dennunzio, E Formenti, L~Manzoni and A E Porreca (Cham: Springer) pp~97--108

\bibitem{DeBaets2018}Bo{\l}t W, Bo{\l}t A, Wolnik B, Baetens J M and De Baets B 2018 A statistical approach to the identification of diploid cellular automata \textit{Theory and Practice of Natural Computing -- TPNC 2017} ed C Mart\'{\i}n-Vide, R~Neruda and M A Vega-Rodr\'{\i}guez (Cham: Springer) pp~37--48

\bibitem{Haye00}Hinrichsen H 2000 Non-equilibrium critical phenomena and phase transitions into absorbing states \textit{Adv. Phys.} \textbf{49} 815--958

\bibitem{jensen}Jensen I 1999 Low-density series expansions for directed percolation: I. A new efficient algorithm with applications to the square lattice \textit{J. Phys. A: Math. Gen.} \textbf{32} 5233--49

\bibitem{munoz}Mu\~{n}oz M A, Dickman R, Vespignani A and Zapperi S 1999 Avalanche and spreading exponents in systems with absorbing states \textit{Phys. Rev.} E \textbf{59} 6175--9

\bibitem{precise}de Mendon\c{c}a J R G 1999 Precise critical exponents for the basic contact process \textit{J. Phys. A: Math. Gen.} \textbf{32} L467--73


\bibitem{Bascompte96}Bascompte J and Sol\'{e} R V 1996 Habitat fragmentation and extinction thresholds in spatially explicit models \textit{J. Anim. Ecol.} \textbf{65} 465--73

\bibitem{Boswell98}Boswell G P, Britton N F and Franks N R 1998 Habitat fragmentation, percolation theory and the conservation of a keystone species \textit{Proc. R. Soc. Lond.} B \textbf{265}~(1409) 1921--25

\bibitem{Szabo05}Oborny B, Mesz\'{e}na G and Szab\'{o} G 2005 Dynamics of populations on the verge of extinction \textit{Oikos} \textbf{109} 291--6

\bibitem{Szabo07}Oborny B, Szab\'{o} G and Mesz\'{e}na G 2007 Survival of species in patchy landscapes: percolation in space and time \textit{Scaling Biodiversity} ed D Storch, P Marquet and J Brown (Cambridge, UK: Cambridge University Press) pp~409--40

\bibitem{Davis08}Davis S, Trapman P, Leirs H, Begon M and Heesterbeek J A P 2008 The abundance threshold for plague as a critical percolation phenomenon \textit{Nature} \textbf{454}~(7204) 634--7

\bibitem{Levin94}Durrett R and Levin S 1994 The importance of being discrete (and spatial) \textit{Theor. Popul. Biol.} \textbf{46} 363--94

\bibitem{Levin96}Levin S A and Durrett R 1996 From individuals to epidemics \textit{Phil. Trans. R. Soc. Lond.} B \textbf{351} 1615--21

\bibitem{Pacala97}Pacala S W and Levin S A 1997 Biologically generated spatial pattern and the coexistence of competing species \textit{Spatial Ecology: The Role of Space in Population Dynamics and Interspecific Interactions} ed D Tilman and P Kareiva (Princeton: Princeton University Press) pp~204--32


\bibitem{Mobilia2007}Mobilia M, Georgiev I T and T\"{a}uber U C 2007 Phase transitions and spatio-temporal fluctuations in stochastic lattice Lotka-Volterra models \textit{J. Stat. Phys.} \textbf{128} 447--83

\bibitem{Uwe2018}Dobramysl U, Mobilia M, Pleimling M and T\"{a}uber U C 2018 Stochastic population dynamics in spatially extended predator-prey systems \textit{J. Phys. A: Math. Theor.} \textbf{51} 063001

\bibitem{Tauber14}T\"{a}uber U C 2014 \textit{Critical Dynamics} (Cambridge, UK: Cambridge University Press)

\bibitem{SzToke98}Szab\'{o} G and T\H{o}ke C 1998 Evolutionary prisoner's dilemma game on a square lattice \textit{Phys. Rev.} E \textbf{58} 69--73

\bibitem{Hauert05}Hauert C and Szab\'{o} G 2005 Game theory and physics \textit{Am. J. Phys.} \textbf{73} 405

\bibitem{Vukov08}Vukov J, Szab\'{o} G and Szolnoki J 2008 Evolutionary prisoner's dilemma game on Newman-Watts networks \textit{Phys. Rev.} E \textbf{77} 026109

\bibitem{janssen}Janssen H K 1981 On the nonequilibrium phase transition in reaction-diffusion systems with an absorbing stationary state \textit{Z. Phys. B} \textbf{42} 151--4

\bibitem{grassberger}Grassberger P 1982 On phase transitions in Schl\"{o}gl's second model, \textit{Z. Phys. B} \textbf{47} 365--74

\bibitem{Takeuchi09}Takeuchi K A, Kuroda M, Chat\'{e} H and Sano M 2007 Directed percolation criticality in turbulent liquid crystals \textit{Phys. Rev. Lett.} \textbf{99} 234503

\item[]Takeuchi K A, Kuroda M, Chat\'{e} H and Sano M 2009 Experimental realization of directed percolation criticality in turbulent liquid crystals \textit{Phys. Rev.} E \textbf{80} 051116

\bibitem{Lemoult16}Lemoult G, Shi L, Avila K, Jalikop S V, Avila M and Hof B 2016 Directed percolation phase transition to sustained turbulence in Couette flow \textit{Nat. Phys.} \textbf{12} 254--8

\end{thebibliography}
\end{document}